\documentclass[modern,twocolumn]{aastex6}
\usepackage{amsmath,amssymb}
\usepackage{aas_macros}
\usepackage{natbib}
\usepackage{graphicx}

\usepackage{ulem,color}

\def\ifnote{\iffalse}

\begin{document}

\title{Determining the Lorentz factor and viewing angle of GRB 170817A}

\author{Yuan-Chuan Zou$^{1*}$,  Fei-Fei Wang$^{1}$, Reetanjali Moharana$^{2}$, Bin Liao$^{1}$, Wei Chen$^{1}$, Qingwen Wu$^{1}$, Wei-Hua Lei$^{1}$, Fa-Yin Wang$^3$}
\affiliation{
$^{1}${School of Physics, Huazhong University of Science and Technology,  Wuhan 430074, China}\\
$^{2}${Racah Institute of Physics, The Hebrew University, Jerusalem 91904, Israel} \\
$^{3}${School of Astronomy and Space Science, Nanjing University, Nanjing 210093, China}\\
 {{*}Email: zouyc@hust.edu.cn}
}

\begin{abstract}
GRB 170817A is a weak short gamma-ray burst (GRB) accompanied by the gravitational wave (GW) event GW170817. It is believed, that an off beaming relativistic jet, produces this weak GRB. Here we use the  $E_{\rm p,i}- E_{\rm iso}$ and $\Gamma - E_{\rm iso}$ relations to determine the Lorentz factor $\Gamma$ and the viewing angle from the edge of the jet  $\theta'_{\rm obs}$ of GRB 170817A. Our result shows, $\Gamma = 13.4 ^{+9.8}_{-5.5}$ and $\theta'_{\rm obs} = 4.3^{+1.8}_{-1.5} \,^{\circ}$. This corresponds to the on-axis $E_{\rm p,i} = 415^{+361}_{-167}$ keV and $E_{\rm iso} = (2.4^{+1.6}_{-1.9}) \times 10^{47}$ ergs of the GRB, which is an intrinsically weak short GRB. Interestingly, the Doppler factor and the luminosity follow a universal relation from GRBs and blazars, which indicates they may share similar radiation mechanism.
\end{abstract}

\keywords{gamma-ray burst: individual; gravitational wave}

\section{Introduction}\label{sec1}
GRB 170817A, was observed by {\it Fermi} GBM \citep{2041-8205-848-2-L14} at 12:41:04.446 UTC as a short gamma-ray bursts (GRBs). Non-detection of this GRB by {\it Insight}-HXMT (Hard X-ray Modulation Telescope) \citep{2017arXiv171006065L}, suggests the burst has very weak fluence and a soft spectrum. The burst is highly noticeable due to its connection to the gravitational wave (GW) GW170817, detected by Laser Interferometer Gravitational-wave Observatory (LIGO) and Virgo, approximately 1.7s before the GBM triggered \citep{PhysRevLett.119.161101}. All the physical quantities of this object is crucial for understanding the event. Comparing to other short GRBs, this has extremely weak luminosity, suggesting the jet to be off-axis to the line of sight \citep[such as][]{2017arXiv170807008J,2017arXiv171005431M,2017arXiv171005438F,2017arXiv171005805W,2017arXiv171005823B,2017arXiv171005839A,2017arXiv171005869H,2017arXiv171005910X,2017arXiv171005931M,2017arXiv171006421G}. The GW fitting parameters suggest the angle to be less than $28^\circ$ \citep{PhysRevLett.119.161101}. Following the fact of no prompt X-ray detection, a number of predictions has been made for the off-axis angle. Observations of $\it Swift$ and $\it NuSTAR$ telescopes, \citet{2017arXiv171005437E} suggests the viewing angle being $\sim 30^\circ$. The constraints from the deep Chandra observation suggests it to be greater than $23^\circ$ \citep{2017arXiv171005852H}. Whereas the radio frequency observational constraints from a relativistic jet, \citet{2017arXiv171005457A} suggest it being $\>20^\circ$. From the upper limit of the ALMA and GMRT at radio bands, \citet{2017arXiv171005847K} found the angle as  $41^\circ$ (or $17^\circ$) leaving other angles still be possible. The modeling of different bands suggested the viewing angle being $\sim 30^\circ$ by \citet{2017arXiv171005905I}, while $\sim 37^\circ - 42^\circ$ by \citet{2017arXiv171006421G}, $25^\circ - 50^\circ$ by \citet{2017arXiv171006426G}. The $E_{\rm p,i}- E_{\rm iso}$ diagram of \citep{2017arXiv171005448P}, shows the GRB 170817A, neither belong to the long GRB nor to the short GRB group, 
supporting the suggestion that GRB 170817A is triggered by an off-axis jet. 
However, the exact angle of the off-axis is still hard to predict due to the fact that, if it is derived from the relativistic beaming effect then it gets degenerated with Lorentz factor of the jet, or if it is derived from the parameter fitting of the GW signal then it gets degenerate with the distance. 

The correlation of different parameters in case of Supernovae and GRBs, have been used in various scenarios. The Philips relation \citep{1993ApJ...413L.105P} is used as the standard candle for cosmology \citep{1997ApJ...483..565P}. The correlations associated with GRBs have also been used to constrain the cosmological parameters \citep[see][for a recent review]{2015NewAR..67....1W}. The relations between redshifts and other quantities have been used to derive the pseudo-redshift of GRBs \citep{2003MNRAS.345..743W}, and the pseudo-redshifts are used to derive the luminosity function \citep{2013ApJ...772L...8T}. The fundamental plane relation \citep{1987ApJ...313...59D} is used to get the peculiar velocity \citep{2001MNRAS.321..277C}.

The host galaxy of GRB 170817A, NGC 4993 is at a nearby distance $40^{+8}_{-14}$ Mpc (redshift $z=0.009783$) \citep{2017arXiv171005444L} compared to other known distance of GRBs. Additionally, this GRB is too weak to be able to detect from far distance. Hence it is reasonable to assume that the far distant GRBs are observed at on-axis.  
Here, we introduce a method in determining the Lorentz factor and viewing angle by using, the peak energy and the isotropic equivalent energy, ( $E_{\rm p,i}- E_{\rm iso}$) \citep{2002A&A...390...81A}, and the Lorentz factor and the isotropic equivalent energy ($\Gamma - E_{\rm iso}$) correlations \citep{2010ApJ...725.2209L} for GRBs like GRB 170817A. 
An independent $E_{\rm p,i}- E_{\rm iso}$ correlation for the short GRBs is not directly available in the literature. Hence we obtained the correlation in section \ref{sec2} with the available data for short GRBs. 
Our method and subsequent result for GRBs 170817A is mentioned in section \ref{sec3}.
Section \ref{sec4} contains a detail discussion and the conclusion derived from our result. 

\section{Relations} \label{sec2}
The widely known Amati relation $E_{\rm p,i}- E_{\rm iso}$, can be used to distinguish between the long GRBs and the short GRBs, as they follow two different relations \citep{2009ApJ...703.1696Z,2013MNRAS.430..163Q}. For the short GRBs, the relation has not been well established due to lack of data. Here, we have collected all the short GRBs with available  $E_{\rm p,i}$ and $E_{\rm iso}$, to check the relation and subsequently in deriving the fitting parameters of the correlation.

The data is very limited as the redshifts of short GRBs are difficult to obtain, compared to long GRBs. Even with measured redshifts, there are still some GRBs, whose spectra can only be fitted with single power-law, hence has no information about $E_{\rm p,i}$, such as GRB 131004A \citep{2013GCN.15316....1S,2013GCN.15315....1X}, GRB 140903A \citep{2014GCN.16768....1P} and GRB 150423A \citep{2015GCN.17740....1U}. 
The data is listed in Tables  \ref{tab:sample1} and  \ref{tab:sample2}. Table  \ref{tab:sample1} contains all the data which are available in the literature, while table \ref{tab:sample2} lists the data, for which we have calculated $E_{\rm iso}$.


\begin{deluxetable*}{cccccc}[b!]
\tablecaption{Short GRBs with $E_{\rm iso}$, collected from literature.  The table lists the name of GRBs, redshifts ($z$), observed duration ($T_{90}$), isotropic equivalent $\gamma$-ray energy ($E_{\rm iso} $) and the spectral peak energy ($E_{\rm p}$).\label{tab:sample1}}
\tablewidth{500pt}
\tabletypesize{\small}
\tablehead{
\colhead{GRB} & \colhead{$z$} & \colhead{$T_{\rm 90}$} & \colhead{$E_{\rm iso} ^{\rm c}$} & \colhead{$E_{\rm p}$} & \colhead{ref $^{\rm d}$} \\
\colhead{} & \colhead{} & \colhead{$(\rm s)$} & \colhead{$10^{52} (\rm ergs)$} & \colhead{$(\rm keV)$} & \colhead{}
}
\startdata
130603B & 0.356 & $0.22^{\rm +0.01}_{\rm -0.01}$ & $0.212^{\rm +0.023}_{\rm -0.023}$ & $660^{\rm +60}_ {\rm -60}$\tablenotemark{b} & (12);(7);(11);(9) \\ 
111117A & $1.31^{\rm +0.46}_{\rm -0.23}$ & $0.576^{\rm +0.143}_{\rm -0.143}$ & $0.34^{\rm +0.13}_{\rm -0.13}$ & $420^{\rm +140}_{\rm -140}$ & (13);(15);(11);(5)  \\ 
101219A & 0.718 & $0.86^{\rm +0.06}_{\rm -0.06}$ & $0.488^{\rm +0.068}_{\rm -0.068}$ & $490^{\rm +103}_{\rm -79}$ & (12);(7);(11);(9)  \\ 
100625A & $0.452^{\rm +0.002}_{\rm -0.002}$ & $0.192^{\rm +0.143}_{\rm -0.143}$ & $0.075^{\rm +0.003}_{\rm -0.003}$ & $486^{\rm +80}_{\rm -80}$ & (13);(15);(11);(9)  \\ 
100206A & 0.408 & $0.128^{\rm +0.091}_{\rm -0.091}$ & $0.0467^{\rm +0.0061}_{\rm -0.0061}$ & $502.8^{\rm +48.9}_{\rm -48.9}$ & (11);(15);(11);(10)  \\ 
100117A & 0.915 & $0.51^{\rm +0.19}_{\rm -0.19}$ & $0.78^{\rm +0.1}_{\rm -0.1}$ & $285.5^{\rm +43.6}_{\rm -43.6}$ & (11);(14);(11);(10)  \\ 
090927 & 1.37 & $0.512^{\rm +0.231}_{\rm -0.231}$ & $0.276^{\rm +0.035}_{\rm -0.035}$ & $59.67^{\rm +1.81}_{\rm -1.81}$\tablenotemark{a} & (11);(15);(11);(16)  \\ 
090510A & 0.903 & $0.96^{\rm +0.138}_{\rm -0.138}$ & $3.95^{\rm +0.21}_{\rm -0.21}$ & $3583^{\rm +468}_{\rm -433}$ & (12);(15);(11);(1)  \\ 
090227B & 1.61 & $1.28^{\rm +1.026}_{\rm -1.026}$ & $28.3^{\rm +1.5}_{\rm -1.5}$ & $1960^{\rm +150}_{\rm -130}$ & (11);(15);(11);(1)  \\ 
081024B & 3.05 & $0.64^{\rm +0.264}_{\rm -0.264}$ & $2.44^{\rm +0.22}_{\rm -0.22}$ & $1405^{\rm +2237}_{\rm -1067}$ & (11);(15);(11);(1)  \\ 
080905A & 0.122 & $0.96^{\rm +0.345}_{\rm -0.345}$ & $0.0658^{\rm +0.0096}_{\rm -0.0096}$ & $586^{\rm +261}_{\rm -110}$\tablenotemark{b} & (11);(15);(11);(14)  \\ 
070729 & 0.8 & $1^{\rm +0.06}_{\rm -0.06}$\tablenotemark{b} & $0.113^{\rm +0.044}_{\rm -0.044}$ & $370^{\rm +375}_{\rm -145}$\tablenotemark{b} & (11);(4);(11);(4)  \\
070429B & 0.904 & $0.32^{\rm +0.006}_{\rm -0.006}$\tablenotemark{b} & $0.0475^{\rm +0.0071}_{\rm -0.0071}$ & $120^{\rm +451}_{\rm -40}$\tablenotemark{b} & (12);(3);(11);(3)  \\ 
061217 & 0.827 & $0.35^{\rm +0.04}_{\rm -0.04}$\tablenotemark{b} & $0.423^{\rm +0.072}_{\rm -0.072}$ & $400^{\rm +490}_{\rm -157}$\tablenotemark{b} & (12);(3);(11);(3)  \\ 
061210A & 0.409 & $0.1^{\rm +0.01}_{\rm -0.01}$\tablenotemark{b} & $0.0024^{\rm +0.0006}_{\rm -0.0006}$ & $540^{\rm +460}_{\rm -187}$\tablenotemark{b} & (11);(3);(11);(3)  \\ 
061201 & 0.111 & $0.86^{\rm +0.03}_{\rm -0.03}$\tablenotemark{b} & $0.0151^{\rm +0.0073}_{\rm -0.0073}$ & $873^{\rm +458}_{\rm -284}$ & (12);(3);(11);(9)  \\ 
060801A & 1.13 & $0.7^{\rm +0.06}_{\rm -0.06}$\tablenotemark{b} & $3.27^{\rm +0.49}_{\rm -0.49}$ & $620^{\rm +647}_{\rm -206}$\tablenotemark{b} & (11);(3);(11);(3)  \\ 
060502B & 0.287 & $0.16^{\rm +0.02}_{\rm -0.02}$\tablenotemark{b} & $0.0433^{\rm +0.0053}_{\rm -0.0053}$ & $340^{\rm +436}_{\rm -115}$\tablenotemark{b} & (12);(3);(11);(3)  \\ 
051221A & 0.547 & $1.24^{\rm +0.02}_{\rm -0.02}$\tablenotemark{b} & $0.263^{\rm +0.033}_{\rm -0.033}$ & $238^{\rm +106}_{\rm -49}$\tablenotemark{b} & (12);(3);(11);(2)  \\ 
000926A & 2.07 & $1.3^{\rm +0.59}_{\rm -0.59}$ & $28.6^{\rm +6.2}_{\rm -6.2}$ & $101^{\rm +6}_{\rm -6}$ & (11);(6);(11);(8)  \\ 
     \hline
\enddata
\tablenotemark{a}{The value in the original paper is in rest-frame, we changed the value using rest-frame divided by $(1+z)$ to observer-frame.}\\
\tablenotemark{b}{The errors in the original papers are with 90$\%$ confidence level, we changed the errors to 1$\sigma$ by multiplying 0.995/1.645.}\\
\tablenotemark{c}{$E_{\rm iso}$ is obtained for rest-frame for $\gamma$-ray energy 1-$10^{\rm 4}$ $\rm keV$ .}\\
\tablenotemark{d}{
References. 
(1) \cite{Ackermann2013};
(2) \cite{Bellm2008};
(3) \cite{Butler2007};
(4) \cite{Butler2010};
(5) \cite{Davanzo2014};
(6) \cite{Frontera2009};
(7) \cite{Golkhou2014};
(8) \cite{Kann2010};
(9) \cite{Li2016ApJS};
(10) \cite{Nava2011};
(11) \cite{Ruffini2016};
(12) \cite{Sang2016};
(13) \cite{Siellez2016};
(14) \cite{Virgili2012};
(15) \cite{Von2014};
(16) \cite{Yu2015B};
}
\end{deluxetable*}

\begin{deluxetable*}{ccccc}[b!]
\tablecaption{Short GRBs, for which $E_{\rm iso}$ is not available in the literature.  
\label{tab:sample2}}
\tablewidth{500pt}
\tabletypesize{\small}
\tablehead{
\colhead{GRB} & \colhead{170428A} & \colhead{160821B} & \colhead{160624A}& \colhead{ref \tablenotemark{d}} 
}
\startdata
$z$ & 0.454 & 0.16 & 0.483 & (4);(5);(2) \\
$D_{\rm L}(10^{\rm 28}~\rm cm)$ \tablenotemark{c} & 0.74 & 0.23 & 0.79 &  \\
$T_{\rm 90}(\rm s)$ & $0.2^{\rm +0.04}_{\rm -0.04}$\tablenotemark{a} & $0.48^{\rm +0.04}_{\rm -0.04}$\tablenotemark{a} & $0.2^{\rm +0.06}_{\rm -0.06}$\tablenotemark{a} & (1);(6);(9) \\
$F_{\rm g}(10^{\rm -6}~\rm ergs~\rm cm^{\rm -2})$ & $4.2^{\rm +0.55}_{\rm -0.54}$\tablenotemark{a} & $1.68^{\rm +0.19}_{\rm -0.19}$ & $0.52^{\rm +0.05}_{\rm -0.05}$ & (8);(7);(3) \\
$\alpha$ & $-0.47^{\rm +0.17}_{\rm -0.13}$\tablenotemark{a} & $-1.37^{\rm +0.22}_{\rm -0.22}$ & $-0.4^{\rm +0.28}_{\rm -0.28}$ & (8);(7);(3) \\
$\beta$ & $-2.46^{\rm +0.31}_{\rm -4.56}$\tablenotemark{a} & ... \tablenotemark{b}& ...\tablenotemark{b} & (8) \\
$E_{\rm p}(\rm keV)$ & $982^{\rm +238}_{\rm -215}$\tablenotemark{a} & $84^{\rm +19}_{\rm -19}$ & $841^{\rm +358}_{\rm -358}$ & (8);(7);(3) \\
$E_{\rm iso}(10^{\rm 52}~\rm ergs~\rm s^{\rm -1})$  & $0.186^{\rm +0.032}_{\rm -0.098}$ & $0.012^{\rm +0.002}_{\rm -0.002}$ & $0.04^{\rm +0.015}_{\rm -0.015}$ & \\
     \hline
\enddata
\tablenotemark{a}{The error in the original paper is in 90$\%$ confidence level, we changed the error into 1$\sigma$ by multiplying 0.995/1.645.}\\
\tablenotemark{b}{The spectral index $\beta$ with ... means the spectrum is best fitted with cutoff power law.}\\
\tablenotemark{c}{$D_{\rm L}$ is calculated using $z$ with the parameters in \cite{Cano2014}, $E_{\rm iso}$ is calculated in 1-10000 keV band using the method in \cite{Schaefer2007}.}\\
\tablenotemark{d}{References. 
(1) \cite{2017GCN.21046....1B};
(2) \cite{2016GCN.19565....1C};
(3) \cite{2016GCN.19570....1H};
(4) \cite{Izzo21059};
(5) \cite{2016GCN.19846....1L};
(6) \cite{2016GCN.19844....1P};
(7) \cite{2016GCN.19843....1S};
(8) \cite{2017GCN.21045....1T};
(9) \cite{2016GCN.19569....1U};}
\end{deluxetable*}

\begin{figure}
\centering
\includegraphics[width=0.5\textwidth]{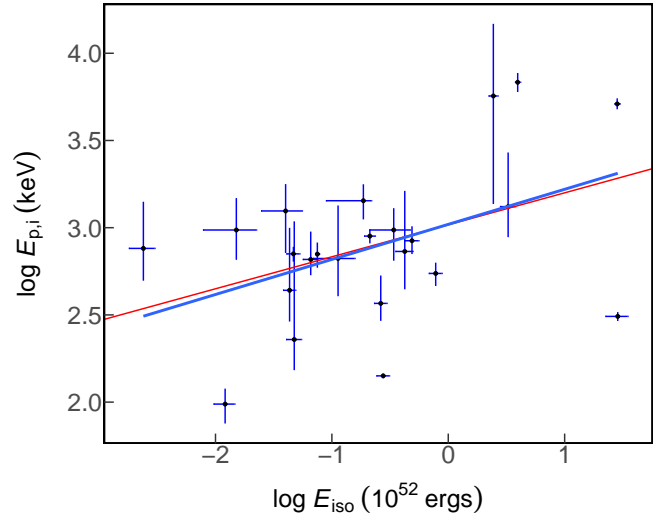}
\caption{The scatter plot for $\log E_{\rm p,i}$ and $\log E_{\rm iso}$. 
The blue line is linear fit with the central values only while 
the red line is the best fit result using the MC simulation.  Data are listed in Tables \ref{tab:sample1} and  \ref{tab:sample2}.
}
\label{fig:epeiso}
\end{figure}

We perform the Pearson's correlation analysis, between $\log E_{\rm iso}$ and $\log E_{\rm p,i}$, with the total 23 short GRBs of Tables \ref{tab:sample1} and  \ref{tab:sample2}. Considering only the central values of the two variables, we obtained Pearson's correlation coefficient, $0.45_{\rm -0.16}^{\rm +0.19}$ with p-value 0.03. 
Further using linear regression fitting method, we obtained the expression as, $\log E_{\rm p,i} = 3.02_{\rm -0.1}^{\rm +0.1} + 0.2_{\rm -0.09}^{\rm +0.09} \times \log E_{\rm iso}$. However the two variables have significant errors associated with observations, and most of the time the errors are asymmetric around the central values. To include the information of errors in the analysis, we used Monte Carlo (MC) simulation. All the errors for our samples are 1$\sigma$. Assuming the errors follow normal distribution, we generated $10^{\rm 4}$ sets of random samples. For the simulated sample, the Pearson's correlation coefficient is $0.31_{\rm -0.12}^{\rm +0.12}$. Using linear regression to the simulated sample we obtained the expression as, $\log E_{\rm p,i} = 3.02_{\rm -0.1}^{\rm +0.1} + 0.18_{\rm -0.08}^{\rm +0.08} \times \log E_{\rm iso}$. The results of the correlation and the data are shown in figure \ref{fig:epeiso}.

Therefore, the $E_{\rm p,i}-E_{\rm iso}$ relation for short GRBs is  
\begin{equation}
  \log E_{\rm p,i} = C_1+ C_2 \log E_{\rm iso,52},
  \label{eqEpEiso}
\end{equation}
with $C_1=3.02 \pm 0.1 $ and $C_2 = 0.18 \pm 0.08$, where $E_{\rm p,i}$ is in unit keV, $E_{\rm iso}$ is in unit of ergs, and $E_{\rm iso,52}=E_{\rm iso}/10^{52}$.



The relation between $\Gamma$ and $E_{\rm iso}$ is taken from \citep{2010ApJ...725.2209L},
\begin{equation}
  \log \Gamma = C_3 + C_4 \log E_{\rm iso,52},
  \label{eqLiang}
\end{equation}
with $C_3=2.26 \pm 0.03$ and $C_4=0.25 \pm 0.03$.

\section{Method and results} \label{sec3}
The off-axis quantities are related to the on-axis quantities following, \citep{1979rpa..book.....R} as,
\begin{equation}
E_{\rm p,off} = a^{-1} E_{\rm p,on}, E_{\rm iso,off} = a^{-3} E_{\rm iso,on},
\label{eqa}
\end{equation}
where $a=\frac{1-\beta \cos \theta'_{\rm obs}}{1-\beta} \simeq 1+(\Gamma \theta'_{\rm obs})^2$ for $\Gamma \gg 1$ and $\theta'_{\rm obs} \ll 1$. $\theta'_{\rm obs} \equiv \theta_{\rm obs}-\theta_{\rm j}$ is the relation between off-viewing angle to the edge of the jet, of which  $\theta_{\rm obs}$ is the viewing angle between line of sight and the jet axis, and $\theta_{\rm j}$ is the opening angle of the jet from the central engine. Here we assume the jet is in top-hat shape for simplicity.

\begin{figure}
\centering
\includegraphics[width=0.5\textwidth]{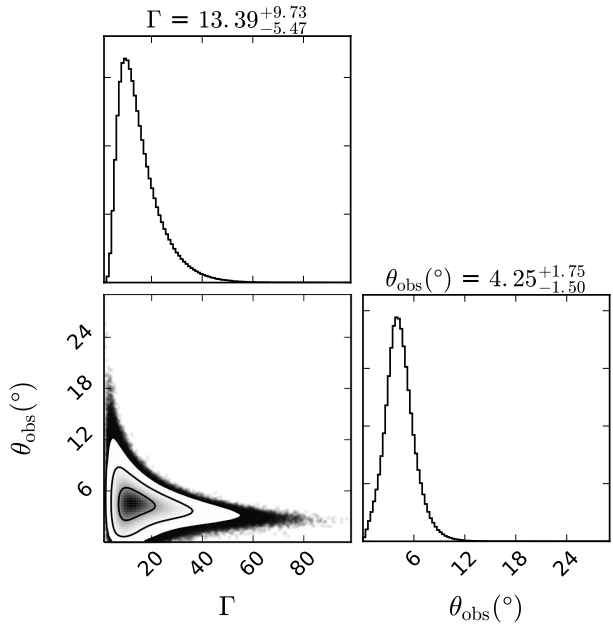}
\caption{The probability distribution of $ \Gamma$ and $\theta'_{\rm obs}$. The lower-left panel shows the scatter plot of the $10^7$ MC simulated samples, with contours $1\sigma, 2\sigma$ and $3\sigma$. The top panel and the right panel are the distributions of $ \Gamma$ and  of $\theta'_{\rm obs}$, respectively. 
This figure is plotted by  a python package -- corner (https://pypi.python.org/pypi/corner).}
\label{figdist}
\end{figure}

The observed $E_{\rm p,obs}$ and $E_{\rm iso,obs}$ of the GRB 170817A, are $(215 \pm 54) {\rm keV}$ and $(2.7\pm0.6) \times 10^{46}$ ergs, respectively \citep{2041-8205-848-2-L14}. Taking the  $E_{\rm p,obs}$ and $E_{\rm iso,obs}$ as  $E_{\rm p,off}$ and $E_{\rm iso,off}$  respectively,  one gets
\begin{equation}
\log \Gamma= C_3 + \frac{C_4}{1-3C_2} (3 C_1 + \log E_{\rm iso,obs,52} - 3 \log E_{\rm p, obs})\label{eqlogGamma}  
\end{equation}
\begin{equation}
\log a = \frac{C_1 + C_2 \log E_{\rm iso,obs,52} - \log E_{\rm p,obs}}{1-3C_2} \label{eqloga}.
\end{equation}
Solving eqs.(\ref{eqlogGamma}) and  (\ref{eqloga}), for the observed values we get $
 \Gamma = 13.4 ^{+9.8}_{-5.5},
\theta'_{\rm obs} = 4.3^{+1.8}_{-1.5} \,^{\circ},
$
and
$\log a=0.30 ^{+0.22}_{-0.28}$. We also get the Doppler factor, $\mathcal{D} = \frac{1}{\Gamma(1-\beta \cos \theta'_{\rm obs})}$ as, $11.8^{+5.3}_{-3.4}$.
The errors are obtained from the MC simulation. The distributions of $\log \Gamma$ and $\theta'_{\rm obs}$ are shown in figure \ref{figdist}.

As the jet opening angle is not known, the viewing angle can not be estimated directly. By taking the jet opening angle roughly as $\theta_{\rm j} \sim 1/\Gamma \sim 6^\circ$  \citep{2007PhR...442..166N}, we get the viewing angle as $10^\circ$. It is well consistent with the estimation of the GW signal \citep{PhysRevLett.119.161101,2017arXiv171005835A}, where the angle is $\leq 28^\circ$ within 90\% confidence limits. Using these physical quantities, we obtained $E_{\rm p,i} = 415^{+361}_{-167}$ keV and $E_{\rm iso} = (2.4^{+1.6}_{-1.9}) \times 10^{47}$ ergs, at the on-axis position. 
Making this GRB, an intrinsically weak GRB, which might be similar to the weak GRB 130603B as suggested by \citet{2017arXiv171005942Y}.

\begin{figure}
\centering
\includegraphics[width=0.5\textwidth]{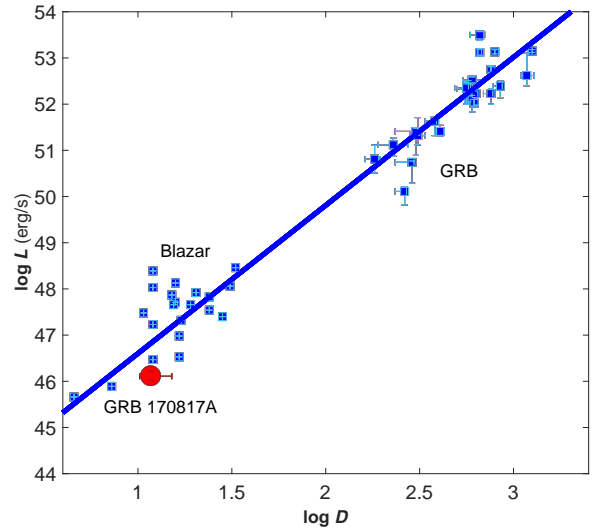}
\caption{The plot of Doppler factor vs isotropic equivalent luminosity for GRBs and blazars using \citet{2011ApJ...740L..21W}. The red dot with errors is GRB 170817A.}
\label{figDL}
\end{figure}

Interestingly, if we plot this GRB to the $\mathcal{D}-L$ relation for both GRBs and blazars \citep{2011ApJ...740L..21W} we found GRB 170817A follows the same slope, as shown in figure \ref{figDL}. This may suggest these different kinds of events share the same radiation mechanism. Note, however this GRB  actually locates in the blazar side.

\section{Discussion and Conclusion}  \label{sec4}

In this work, we derived the $E_{\rm p,i}- E_{\rm iso}$ relation for short GRBs using the latest observations. Using the result and the $\Gamma - E_{\rm iso}$ correlation, we calculated the Lorentz factor as, $\Gamma = 13.4 ^{+9.8}_{-5.5}$ and the viewing angle as $\theta'_{\rm obs} = 4.3^{+1.8}_{-1.5} \,^{\circ}$ following the top-hat jet edge. These values can also give the on-axis, $E_{\rm p,on,i} =  415^{+361}_{-167}$ keV and $E_{\rm iso,on}=(2.4^{+1.6}_{-1.9}) \times 10^{47}$ ergs, suggesting the GW170817 associated GRB 170817A is a relatively weak short GRB. The estimated viewing angle from the jet axis is around $10^\circ$, which is well consistent with the inclination angle inferred from the GW signal.

Here we used the $\Gamma- E_{\rm iso, on}$ relation for both long and short GRBs. 
One can also engage other relations for independent estimations, such as using the $\Gamma-L_{\rm iso}$ relation \citep{2012ApJ...751...49L,2012ApJ...755L...6F} or the $L_{\rm iso} - E_{\rm p,i} - \Gamma$ relation \citep{2015ApJ...813..116L} to replace the $\Gamma-E_{\rm iso}$ relation,  and using the $E_{\rm p,i} - L_{\rm iso}$ relation \citep{2003MNRAS.345..743W,2004ApJ...609..935Y} to replace the $E_{\rm p,i} - E_{\rm iso}$ relation.


For simplicity we used $\theta'_{\rm obs}=\theta_{\rm obs}-\theta_{\rm j}$ for the off-axis frame transformation. However a careful calculation will include the effect of different angles within $\theta_{\rm j}$ and the corresponding arrival time.
Again we haven't considered the structured jet as mentioned in \citep{2017arXiv171005910X}. Consideration of structured jet would 
require involvement of 
different models and distinguishing the suitable model for this GRB, which it may not be decisive.
However, $\theta'_{\rm obs}$ can be taken as an effective off axis angle if the $\theta_{\rm j}$ and the jet structure are considered.

The independent method of determining the viewing angle, can be used to reduce the distance uncertainty directly from the GW, as the distance and the angle is highly coupled. Therefore, in the future if electromagnetic counterpart observed with GW event, one can get a more precise distance directly from the GW signal. With the precise distance measurement, which is an independent standard siren, the cosmology parameters (especially $H_0$) can be obtained with high confidence.

The observed GRB 170817A is known to be from NS-NS merger, and it is a weak short GRB. It is possible that the NS-NS merger may produce a weak short GRB, while the BH-NS
merger produces a strong short GRB. As we have seen from other GWs, such as GW 150914 \citep{2016PhRvL.116f1102A}, with tens of solar masses. Therefore, the most common stellar BH might have
mass tens of solar mass. Because the GRB energy is intrinsically from the
gravitational energy released, the more massive BH-NS mergers are
naturally connected to the stronger GRBs, while NS-NS mergers produce weak ones. Consequently, the short GRBs may be classified into two subclasses. 

\section*{Acknowledgments}
We thank the helpful discussion with Yizhong Fan, Enwei Liang and Shuaibing Ma. This work is supported by the National Basic Research Program of China (973 Program, Grant No. 2014CB845800) and by the National Natural Science Foundation of China (Grants No. 11773010, U1231101  and U1431124).


\end{document}